\documentclass{aa}
\usepackage{aabib99}
\usepackage{psfig}

\begin{document}
   \thesaurus{2         
              (11.03.1;  
               11.03.4 Cl0024+17;  
               12.03.3;  
               12.04.1;  
               12.07.1;  
               13.25.2)} 

\title{X-ray observations and mass determinations in the cluster of galaxies 
Cl0024+17
\thanks{Based on observations with the ROSAT and ASCA satellites  
and with the Canada-France-Hawaii Telescope at Mauna Kea, Hawaii, USA.}
}

\author{
G. Soucail\inst{1} 
\and N. Ota\inst{2}
\and H. B\"ohringer \inst{3} 
\and O. Czoske\inst{1} 
\and M. Hattori \inst{4} 
\and Y. Mellier \inst{5,6}
   }

   \offprints{G. Soucail, soucail@obs-mip.fr}

   \institute{Observatoire Midi-Pyr\'en\'ees, Laboratoire d'Astrophysique, 
     UMR 5572, 14 Avenue E. Belin, F-31400 Toulouse, France 
   \and Institute of Space and Astronautical Science, 3-1-1
     Yoshinodai, Sagamihara, Kanagawa 229-8510, Japan
   \and Max-Planck-Institut f\"ur Extraterrestrische Physik, 
     Gie\ss enbachstr., D-85740 Garching bei M\"{u}nchen, Germany
   \and Astronomical Institute, T\^ohoku University, Aoba Aramaki,
     Sendai 980, Japan 
   \and Institut d'Astrophysique de Paris, 98 bis Boulevard Arago, F-75014 
     Paris, France
   \and Observatoire de Paris, DEMIRM, 61 Avenue de l'Observatoire, F-75014
     Paris, France
   }
   
\date{Received November 4, 1999, accepted January 18, 2000}

\maketitle

\markboth{G. Soucail et al.:Mass determinations in Cl0024+17}{}

\begin{abstract} 
We present a detailed analysis of the mass distribution in the rich
and distant cluster of galaxies Cl0024+17. X-ray data come from both a
deep ROSAT HRI image of the field \cite{bohringer99} and ASCA spectral
data. Using a wide field CCD image of the cluster, we optically
identify all the faint X-ray sources, whose counts are compatible with
deep X-ray number counts. In addition we marginally detect the X-ray
counter-part of the gravitational shear perturbation detected by
Bonnet et al. \cite*{bonnet94} at a 2.5 $\sigma$ level. A careful
spectral analysis of ASCA data is also presented. In particular, we
extract a low resolution spectrum of the cluster free from the
contamination by a nearby point source located 1.2 arcmin from the
center. The X-ray temperature deduced from this analysis is $T_X = 5.7
^{+4.9}_{-2.1}$ keV at the 90\%\ confidence level. The comparison
between the mass derived from a standard X-ray analysis and from other
methods such as the Virial Theorem or the gravitational lensing effect
lead to a mass discrepancy of a factor 1.5 to 3. We discuss all the
possible sources of uncertainties in each method of mass determination
and give some indications on the way to reduce them. A complementary
study of optical data is in progress and may solve the X-ray/optical
discrepancy through a better understanding of the dynamics of the
cluster.

\end{abstract}

\section{Introduction} 
Clusters of galaxies are the most massive gravitationally bound
systems in the Universe. They are dynamically young as most of their
time scales for evolution (virialization, cooling of the intra-cluster
gas, dynamical friction, two-body relaxation, \ldots) are not small
with respect to the age of the Universe. One important cosmological
issue is to understand the distribution and the physical properties of
the different components inside clusters of galaxies. The study of
each component can allow an estimate of the total mass and its
distribution through 3 independent methods of mass determination.  The
first and oldest one is the Virial Theorem which relates the galaxy
distribution to the total mass \cite{zwicky33}. It is a global mass
estimator which is mainly uncertain due to departures from dynamical
equilibrium such as substructure and infall
\cite{geller82,merrit87,merrit94}.  More recently, progress in X-ray
astronomy has stressed the importance of the intra-cluster gas both as
a major component of clusters \cite{sarazin86} and as a good tracer of
the mass under the assumptions of spherical symmetry of the
distributions and thermal equilibrium of the X-ray emitting gas.
Another very promising mass estimator is the mass deduced from
gravitational lensing. Two regimes of lensing are to be considered in
rich clusters. The first one is the strong lensing regime, where
multiple images can be formed (see Fort \& Mellier \cite*{fort94} for
a review). Lens modeling of these multiple images gives strong
constraints on the mass, but only the central projected mass can be
estimated that way \cite{mellier93,kneib96}.  For the weak lensing
regime, the shear-to-mass inversion is still rather uncertain,
although recent theoretical progress opens new perspectives on the
quantitative estimate of the deflecting masses
\cite{seitz98,mellier99}.

Cl0024+17 ($z = 0.39$) is one of the few lensing clusters for which a
combined analysis of its mass distribution is now possible. It was
initially studied by Butcher and Oemler \cite*{butcher78} and was
pointed out for its high content of blue cluster members. It is a rich
cluster, highly concentrated in the center but not dominated by a cD
galaxy. On the contrary, there is a concentration of bright
galaxies. Dressler et al. \cite*{dressler85} obtained a reasonable
number of redshifts of cluster members, from which they measured a
rest-frame velocity dispersion of 1300 km s$^{-1}$ \cite{schneider86}.
The spectacular system of giant arcs in the center of the cluster was
initially mentioned by Koo \cite*{koo88} and observed
spectroscopically by Mellier et al. \cite*{mellier91} with no clear
redshift determination. A tentative measurement of $z = 1.675$ has
recently been claimed by Broadhurst et al.  \cite*{broadhurst99} for
this very blue multiple system.  Deep HST images revealed that this
arc system corresponds to the merging of three images, with the clear
identification of an additional counter-image
\cite{kassiola92,smail96,colley96}.  Several arclets are also present
in the same region.  In addition to the strong lensing effect detected
in the core of the cluster, a significant weak shear signal was
measured by Bonnet et al. \cite*{bonnet94} up to 3 $h_{50}^{-1}$ Mpc
from the center. This was indeed the first distortion map measured on
a massive cluster and the shear was detected up to a 10\%\ level at
the periphery. Finally, Cl0024+17 was detected by the Einstein
Observatory with an X-ray luminosity of $2.7 \times 10^{44} \
h_{50}^{-2}$ erg s$^{-1}$ \cite{henry82}.

In this paper, we analyze X-ray data obtained with both the ROSAT HRI
and with ASCA. Section 2 presents the X-ray data, while in Section 3,
the spatial analysis of the deep ROSAT HRI image gives quantitative
measures for the different sources detected in the field, in
particular for the cluster emission. Section 4 deals with the spectral
analysis of the ASCA data and Section 5 proposes a detailed evaluation
of the different mass determinations available for this cluster.  In
Section 6 we discuss the different sources of uncertainties in mass
estimates, which may be related to the strong mass discrepancy
identified in this cluster. Section 7 summarizes the main results of
this paper.

Throughout the paper, we use a Hubble constant of H$_0 = 50 h_{50}$ km
s$^{-1}$ Mpc$^{-1}$, with $\Lambda = 0$ and $\Omega_0 = 1$. At the cluster
redshift ($z=0.39$), 1\arcsec\ corresponds to 6.36 $h_{50}^{-1}$ kpc.

\section{X-ray data}
\subsection{ROSAT HRI data}
\begin{figure*}
\centerline{\psfig{figure=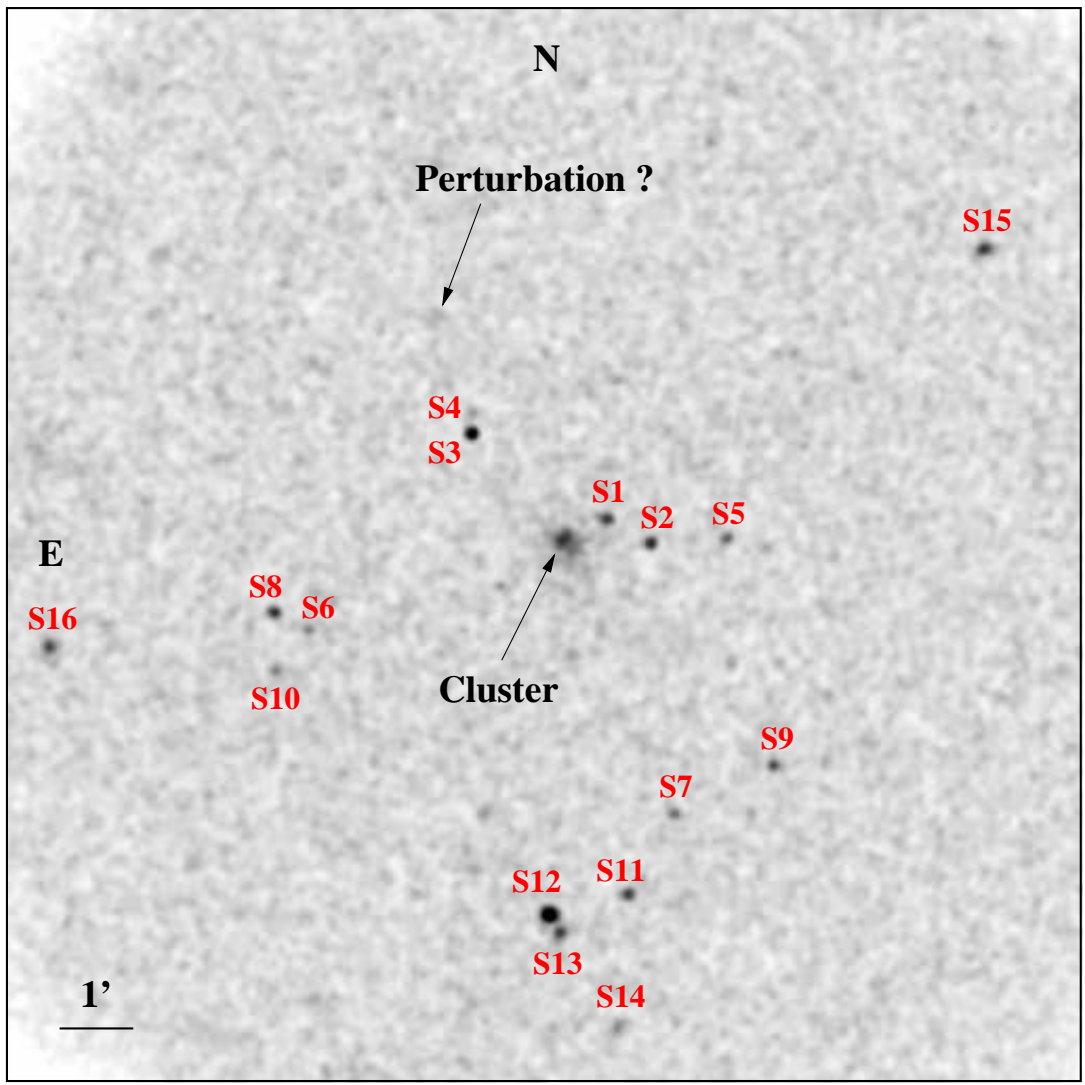,angle=-90,width=0.8\textwidth}}
\caption{HRI image of the field of Cl0024+17 with the identification
of most of the detected individual sources. The size of the image
corresponds to 28\arcmin $\times$ 28\arcmin . North is up, East is
left.
Data have been smoothed with a Gaussian filter ($\sigma = 6''$
or 14 \arcsec\ FWHM, see text for details).}
\label{hri_map}
\end{figure*}

Cl0024+17 was observed by the ROSAT HRI between January 1994 and July
1996 for a total integration time of 116550 sec. More details about
the data reduction of the image can be found in B\"ohringer et al.
\cite*{bohringer99}. For our purpose the data were smoothed with a
Gaussian filter ($\sigma = 6 ''$ or equivalently 14\arcsec\
FWHM). This value is slightly higher than the resolution of the HRI
but it is a good compromise with the low signal level of the
image. The final detection level on the smoothed image corresponds to
$3\sigma = 2.6 \times 10^{-7}$ cts s$^{-1}$arcsec$^{-2}$ and the
measured background level is $1.1 \times 10^{-6}$ cts
s$^{-1}$arcsec$^{-2}$ in good agreement with most of the data from the
HRI \cite{david95}.  Figure \ref{hri_map} displays a wide part of the
HRI image of the field. The cluster extended emission is clearly
detected, although rather faint and not prominent in the
image. Several sources are also detected and discussed below (see
Section 3.2).

\begin{figure}
\psfig{figure=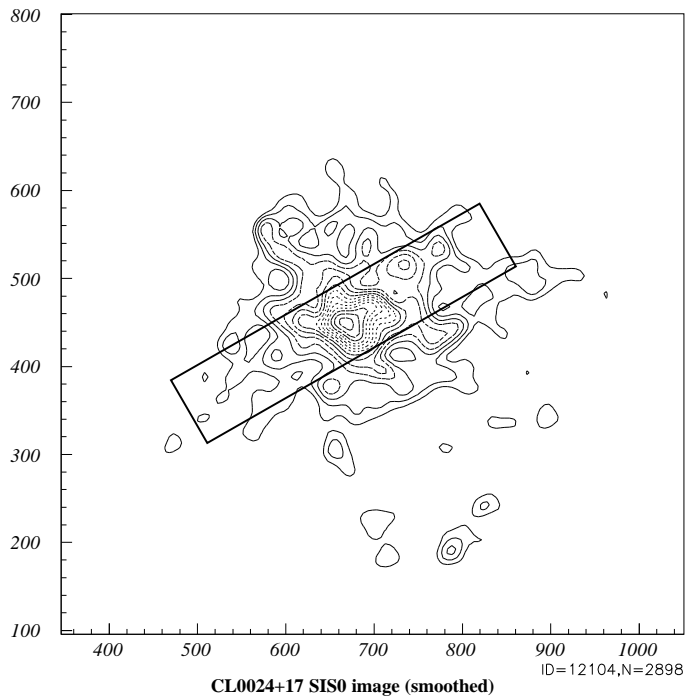,angle=0,width=0.5\textwidth}
\caption[]{X-ray image of the field of Cl0024+17 obtained with the
ASCA SIS. The one-dimensional projection position is shown with a
rectangle. North is up and East is left. The pixel size is 0.026
arcmin and the image size is $18.6' \times 18.6'$.}
\label{sis_image}
\end{figure}

\subsection{ASCA data}
Cl0024+17 was observed with the ASCA GIS and SIS on July 21,
1996. X-ray emission centered at $\alpha_{2000}= 0\fh 26 \fm 39 \fs$,
$\delta_{2000}= 17\fd 09\farcm 45\farcs$ is detected, with an error
circle of 1 arcmin due to the uncertainty of absolute attitude
determinations of ASCA. This is consistent with the cataloged position
of Cl0024+17 \cite{henry82}. There are several sources in the vicinity
of the cluster in the SIS field (Figure \ref{sis_image}). In this
paper, we only use the SIS data because the nearby sources were not
spatially resolved with the GIS. The total effective observation time
after data filtering is 46 ksec for the SIS.  The intrinsic cluster
count rate is $(9.1 \pm 0.7)\times10^{-3}$ cts s$^{-1}$ for the SIS
within the cluster region defined in Section 4.1.

\section{Spatial analysis of the X-ray emission}
\subsection{X-ray/optical centering}
The coordinates of ROSAT in the center of the HRI image were given as
$\alpha_{2000} = 0 \fh 26 \fm 33 \fs $ and $\delta_{2000} = 17 \fd 09
\farcm 36 \farcs $. The accuracy in the absolute pointing of the
telescope is better than 10\arcsec\ but still too large for a correct
identification of the optical counter-parts of the X-ray sources.  The
optical centering was done with the help of a wide field mosaic of $3
\times 3$ images taken at the ESO NTT in a 24\arcmin $\times$
24\arcmin\ field in total. Each frame is a 10 minute exposure, with a
pixel size of $0\farcs7$. This kind of image is not very useful for a
morphological analysis of the individual galaxies or for any weak
shear measurement, but it can help with the optical identification of
the X-ray sources.  In the field, more than 20 stars selected from the
Guide Star Catalog and not saturated on the CCD image were
identified. The correspondence between their equatorial coordinates
and their position on the CCD was obtained by a linear regression
and the residuals were limited to $0\farcs2$ typically. No rotation
was necessary, as both images were correctly aligned along the
North direction. The center of the X-ray image was then positioned
with its own coordinates on the CCD image with a correct
scaling. Another wide field image was used, taken with the UH8K
camera at CFHT. It is a 20 minutes exposure in I, observed in very
good seeing conditions ($0\farcs6$)and covering a field of view of
28\arcmin $\times$ 28\arcmin . Similar procedures for the
astrometry were followed, using stars identified in the APM
catalogue ({\tt http://www.ast.cam.ac.uk/\~\,apmcat
}), resulting in a non-linear astrometric solution for
each CCD chip of the UH8K camera.

\subsection{Identification of individual sources}
\begin{table*}
\caption[]{Identification of the sources detected in the HRI image.
They are listed with increasing distance from the cluster center. CR
is the count rate in the HRI image. The apparent unabsorbed flux (in
erg s$^{-1}$ cm$^{-2}$) was computed assuming a power-law spectrum
with spectral index of 1 for the sources and hydrogen absorption with
$N_H = 4.2 \times 10^{20}$ cm$^{-2}$, with a conversion factor $3.35
\times 10^{-11}$ erg s$^{-1}$ cm$^{-2}$ for a CR of 1.  }
\label{tab-sources}
\begin{flushleft}
\begin{tabular}{lcccccccl}
\hline\noalign{\smallskip}
 & $\alpha_{2000}$ & $\delta_{2000}$ & CR (s$^{-1}$)& Flux [0.5--2 keV] 
 & V & I & V--I & Comments \\
\noalign{\smallskip}
\hline\noalign{\smallskip}
S1  & 00:26:31.3 & 17:10:16.2 & $8.7 \ 10^{-4}$ & $2.9 \ 10^{-14}$ 
 & 20.56 & 18.84 & 1.72 & $z = 0.4017$ \\
 & & & &
 & 19.56 & 18.08 & 1.48 & $z = 0.2132$ \\
S2  & 00:26:26.4 & 17:09:38.1 & $7.4 \ 10^{-4}$ & $2.5 \ 10^{-14}$
 & 20.01 & 19.42 & 0.59 & PC 0023+1653, $z=0.959$ \\
S3  & 00:26:46.2 & 17:12:31.6 & $1.4 \ 10^{-3}$ & $4.7 \ 10^{-14}$
 & 20.31 & 19.36 & 0.95 & \\
S4  & 00:26:46.1 & 17:13:05.2 & $3.6 \ 10^{-4}$ & $1.2 \ 10^{-14}$
 & 21.59 & 20.65 & 0.94 & \\
S5  & 00:26:18.1 & 17:09:46.3 & $6.0 \ 10^{-4}$ & $2.0 \ 10^{-14}$
 & --- & 17.88 & --- & not separated in V \\
S6  & 00:27:04.1 & 17:07:20.6 & $4.9 \ 10^{-4}$ & $1.6 \ 10^{-14}$
 & --- & --- & --- & not separated in V, gap in I \\
S7  & 00:26:23.8 & 17:02:28.8 & $5.4 \ 10^{-4}$ & $1.8 \ 10^{-14}$
 & 21.28 & 19.80 & 1.48 & \\
S8  & 00:27:07.9 & 17:07:48.5 & $7.0 \ 10^{-4}$ & $2.3 \ 10^{-14}$
 & 21.45 & 20.44 & 1.01 & \\
S9  & 00:26:13.0 & 17:03:46.5 & $6.4 \ 10^{-4}$ & $2.1 \ 10^{-14}$
 & 21.87 & 20.15 & 1.72 & \\
S10 & 00:27:07.8 & 17:06:17.1 & $4.5 \ 10^{-4}$ & $1.5 \ 10^{-14}$
 & 20.22 & 19.68 & 0.54 & \\
S11 & 00:26:29.0 & 17:00:22.6 & $8.2 \ 10^{-4}$ & $2.7 \ 10^{-14}$
 & 21.06 & 19.82 & 1.24 & \\
S12 & 00:26:37.6 & 16:59:50.5 & $2.9 \ 10^{-3}$ & $9.7 \ 10^{-14}$
 & 20.40 & 19.06 & 1.34 & 2E 0024.0+1643 \\
S13 & 00:26:36.5 & 16:59:22.0 & $1.0 \ 10^{-3}$ & $3.4 \ 10^{-14}$
 & 20.94 & 19.28 & 1.66 & $z=0.4083$ \\
S14 & 00:26:30.1 & 16:56:52.9 & $7.4 \ 10^{-4}$ & $2.5 \ 10^{-14}$
 & --- & 19.75 & --- & outside V field \\
S15 & 00:25:49.7 & 17:17:23.0 & $1.2 \ 10^{-3}$ & $4.0 \ 10^{-14}$
 & 20.55 & 19.47 & 1.08 & 2E 0023.2+1700 \\
S16 & 00:27:32.7 & 17:06:53.5 & $1.2 \ 10^{-3}$ & $4.1 \ 10^{-14}$
 & --- & 18.42 & --- & outside V field \\
\noalign{\smallskip}
\hline
\end{tabular}
\end{flushleft}
\end{table*}

\begin{figure*}
\centerline{\psfig{figure=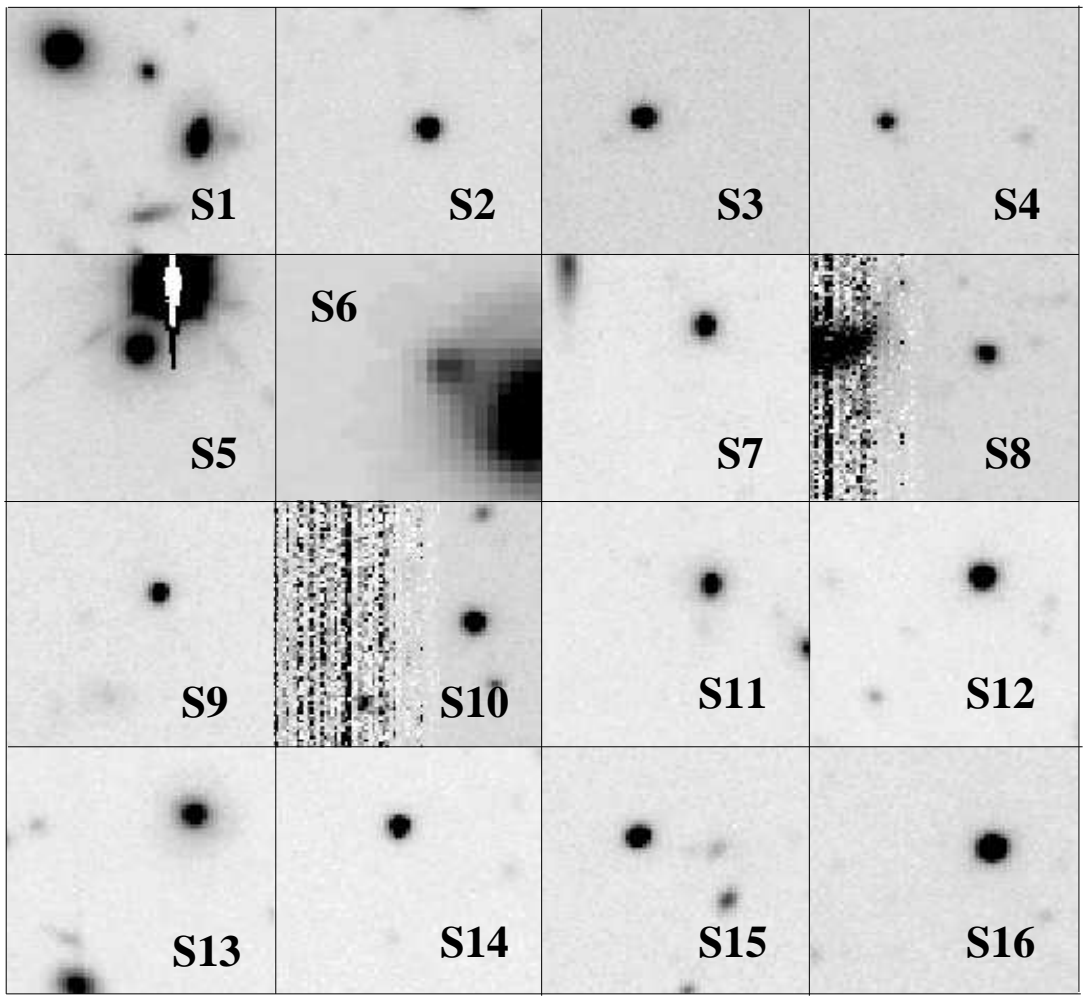,width=0.8\textwidth}}
\caption{Optical identification of the X-ray individual sources in the
field of Cl0024+17. Each sub-image corresponds to a 16\arcsec\ field
centered on the X-ray centroid. North is up, East is left. Except for
S6 the sub-images are taken from the UH8K I-band image, with a pixel
size of $0\farcs205$. S6 is located in the gap between two adjacent
CCD chips so we used the NTT V sub-image in that case, with a larger pixel
sixe of $0\farcs7$. See more comments in the text. }
\label{fig-sources}
\end{figure*}

The first goal after the X-ray/optical centering was to search for a
possible X-ray counter-part of the weak shear perturbation detected by
Bonnet et al. \cite*{bonnet94}. This perturbation was claimed to be a
massive structure which might be a cluster substructure or a chance
superposition of a group of galaxies at a different redshift.  In the
X-ray image, a weak over-density is seen close to the location of the
shear perturbation, with a maximum detected at a 5$\sigma$ level above
the background. The total number counts is around $35 \pm 14$ photons
above the background in the image, or a total count rate of $(3 \pm 1.2)
\times 10^{-4}$ s$^{-1}$ above the background. It is a poor detection
(extended low surface brigthness object) which requires deeper data
to confirm whether it is real or not. In the case this X-ray emission
is at the cluster redshift, we can roughly estimate its luminosity,
using the conversion factor discussed in Section 4.2 and a
temperature of 1 keV, giving a luminosity in the ROSAT band of $1
\times 10^{43} h_{50}^{-2}$ erg s$^{-1}$, typical of group
emission. A similar result was found by Erben et al. \cite*{erben99}
in the cluster Abell 1942: a so-called ``dark clump'' inducing a
significant perturbation of the shear field may have an X-ray
counter-part with similar luminosity, if indeed this perturbation is
at the same redshift as the cluster.

In addition, several other sources are detected in the field within a
radius of 14\arcmin\ around the cluster center (which corresponds to
the unvignetted field of the HRI). We limited our sample to a total
of 40 counts on the image, corresponding to a count rate of $3.4
\times 10^{-4}$ s$^{-1}$ or a flux limit of $1.2 \times 10^{-14}$ erg
s$^{-1}$ cm$^{-2}$ with the assumptions for the input spectra 
specified in the caption of Table \ref{tab-sources}. They are labeled
S1 to S16 with increasing distance from the cluster center and their
properties are summarized in Table
\ref{tab-sources}. All these sources are compatible with
point-like sources, although for the faintest ones this is less
clear. For each source, we searched for an optical counter-part in
a 8\arcsec\ radius circle which corresponds to the typical ROSAT
error box (Figure
\ref{fig-sources}).  The source S2 was immediately identified with
the quasar J002626.2+170937
\cite{veron98}, initially observed by Schmidt et
al. \cite*{schmidt86}. This object is also labeled PC 0023+1653 and is
at redshift $z = 0.959$. This is the most secure identification in
this X-ray map, so we decided to make this source exactly coincident
with its optical counter-part for a final X-ray/optical centering. Its
luminosity, computed assuming a power-law index for its spectrum is
$L_X (\rm S2) = 1.4 \times 10^{44} h_{50}^{-2}$ erg s$^{-1}$ in the
[0.5-2] keV band, a typical value for such an object. For most of
the sources, a relatively bright point-like object can be identified
within the ROSAT error box ($V \sim 20-21, I
\sim 18-19$). In the case of S1, there is an uncertainty in the optical
identification. From a spectroscopic program dedicated to a wide
field redshift survey in the cluster
\cite{czoske00} we have got the spectra of the two brightest objects
located in the ROSAT error box. The most central one is a typical
cluster member ($z=0.4017$) while the second, more distant and
brighter object, is a foreground star-forming galaxy at
$z=0.2132$. None of the spectra present strong signs of nuclear
emission, although the foreground object is an emission line galaxy,
but with only weak forbidden lines. For S5 and S6 there is a
point-source object close to a much brighter source (a star for S5 and
a spiral galaxy for S6). We also mention S13 which is clearly an
extended object at redshift $z=0.4083$ showing an extended disc viewed
face-on. Its spectrum shows absorption lines and no signs of nuclear
activity. Unfortunately no spectra are available for the other objects
and the spectroscopic identification remains to be done.

We also checked if any of these sources could be identified with
previously know X-ray sources. Indeed, S12 and S15 correspond to
sources detected with the HEAO-2 Einstein satellite
\cite{mcdowell94} although their positions differ by 10 to 20\arcsec
. But this can be due to uncertainties in the distortion correction in
the ROSAT image for these off-center sources or to uncertainties in
the previous localization of these sources.

Finally, the number counts of X-ray sources (excluding the cluster) in
our HRI image can be compared to the expected number, for a given flux
limit. For $S > 1.2 \times 10^{-14}$ erg s$^{-1}$ cm$^{-2}$, we find
16 sources within 600 sq. arcmin, and 12 sources with $S > 2 \times
10^{-14}$ erg s$^{-1}$ cm$^{-2}$. These values are compatible
with the counts observed in deep surveys \cite{hasinger98} with a
slight excess not exceeding a $1.5 \sigma$ level. 

\subsection{The cluster emission} 
\begin{figure*}
\centerline{
\psfig{figure=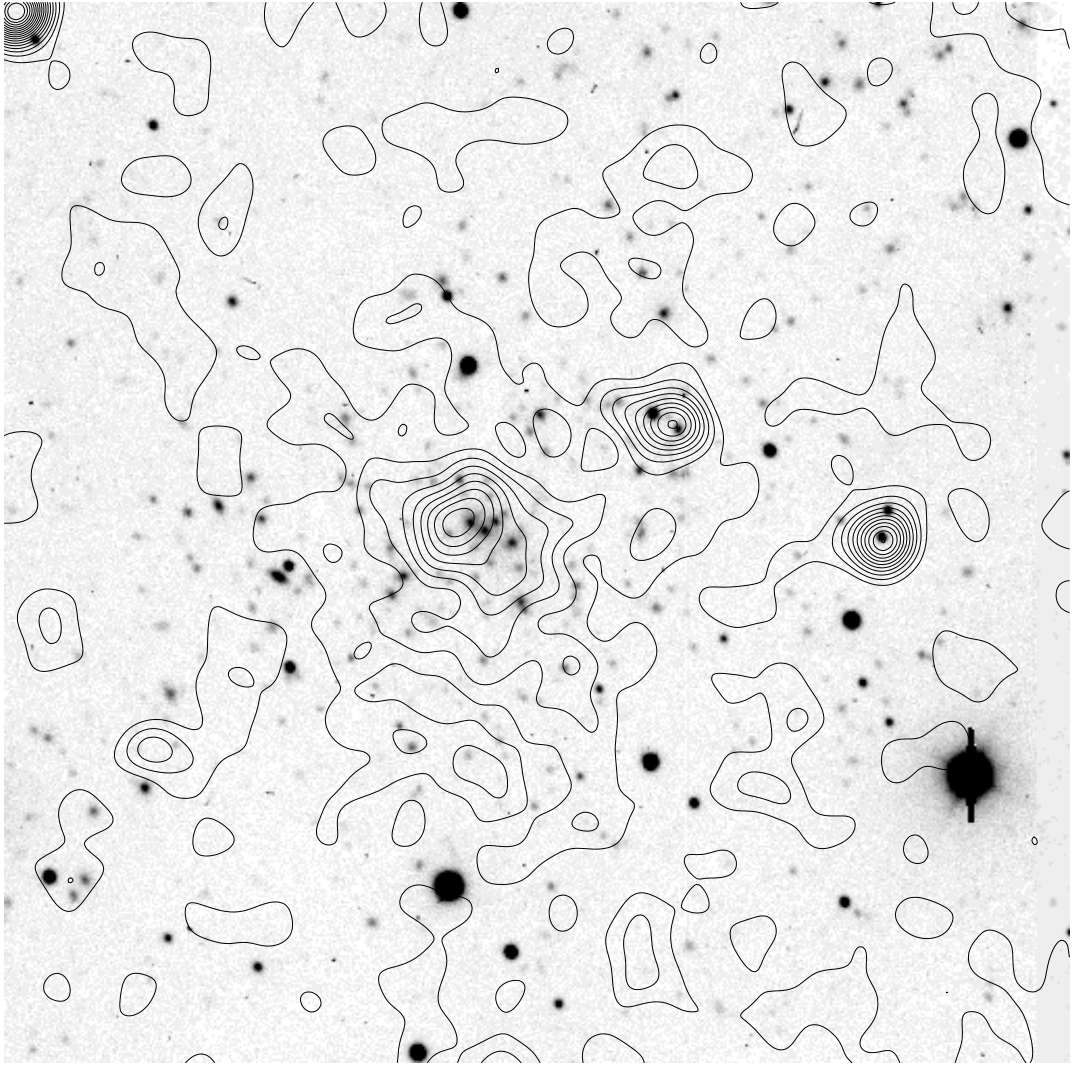,width=0.8\textwidth}
}
\caption{CCD image of the center of the cluster Cl0024+17 ($6' \times
6'$). The X-ray contours are overlayed, with a linear spacing. First
contour corresponds to 1 $\sigma$ above the background and the step is 
approximatively 3 $\sigma$. North is up, East is left.}
\label{opt_X}
\end{figure*}

\begin{figure}
\psfig{figure=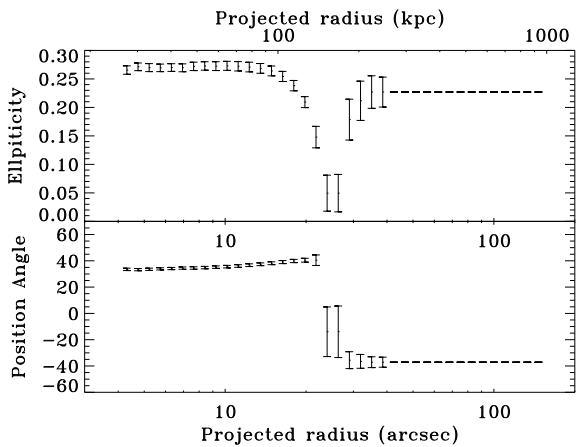,width=0.5\textwidth}
\caption{Plot of the ellipticity and the position angle of the 
semi-major axis versus radius, from the elliptical fit of the 
X-ray isophotes. The origin of the PA corresponds to the North direction, 
and positive angles correspond to a clockwise rotation of the major
axis.}
\label{ellip_pa}
\end{figure}

With the correct matching between X-ray map and optical CCD image, we
first verified that the X-ray cluster center coincides roughly with
the location of the central galaxies (Figure
\ref{opt_X}). In addition, the cluster X-ray emission clearly displays
an elliptical shape, so we proceeded to an elliptical fit of the isophotes,
with the ELLIPSE package in the IRAF/STSDAS environment. In order to avoid
contamination by the individual sources, S1 and S2 were masked. Because
of the small number of photons detected from the cluster contribution, a
good elliptical fit was not possible far from the center and we let the
ellipse parameters fixed at radius larger than 30\arcsec\ with the values
found at 30\arcsec . In the inner part there is a significant twist of the
isophotes, with a change in the PA of nearly 90 degrees (Figure
\ref{ellip_pa}). The inner ellipticity is rather constant
($\varepsilon_{in} \simeq 0.3$) while the outer one is fixed to
$\varepsilon_{out} = 0.25$, with $\varepsilon = 1 - b/a$ in both
cases. The main axis orientation is 40\degr\ clockwise from the North
direction in the center and changes to 40\degr\ counter-clockwise in
the outer part of the cluster. There is also a significant shift of
the X-ray centroid, of about 12\arcsec\ with respect to the centroid
of the outer ellipses. This may be an indication of some physical
processes occurring in the cluster center which deviates from a
regular isothermal distribution of the gas. Anyway, the outer
elliptical shape of the X-ray isophotes is quite similar to the
isoluminosity distribution of the cluster galaxies
\cite{czoske00}, with an ellipticity of about 0.3 and a position
angle of --45\degr\ from the North direction.  The X-ray signal
from the cluster contribution is detected up to 150\arcsec\ from
the center, which corresponds to a 1 $h_{50}^{-1}$ Mpc radius at
the cluster redshift.

The fit of the surface brightness profile was done in two ways. First,
we used the radial intensity profile extracted from the elliptical fit
and second, we extracted a profile inside circular annuli, which better
corresponds to the use of the standard de-projection techniques. In
both cases we fitted the profile with the ``classical'' $\beta$-profile
\cite{cavaliere76,jones84}. Our results are quite consistent with those
found by B\"ohringer et al. \cite*{bohringer99} who use a more refined
technique including de-convolution by the ROSAT HRI PSF, except in one
point: in order to avoid the un-physical meaning of a low $\beta$
value, we limited the fit to $\beta$ larger than 0.5, and consequently
found a larger value for the core radius $r_c$. But in the rest of the
paper, we will use their values of the X-ray profile which seem quite
reliable, especially due to the correction for the instrumental PSF:
\begin{eqnarray}
S_0 & = & 4.17 \times 10^{-6} \ \hbox{cts s}^{-1} \hbox{arcsec}^{-2} 
\nonumber \\
r_c & = & 10\farcs4 ^{+6.1}_{-3.9} \ = 66 ^{+38}_{-25} \ h_{50}^{-1} \ 
\hbox{kpc} \nonumber \\
\beta & = & 0.475^{+0.075}_{-0.05} \nonumber
\end{eqnarray}

\section{Spectral analysis}
\subsection{ASCA observations}
Since the angular separation between the centre of Cl0024+17 and S1 is
only 1.2 arcminutes, the contamination of the Cl0024+17 spectrum by
the source S1 should be carefully treated.  In order to obtain the
intrinsic cluster spectrum, we performed 1-dimensional image fitting
and separated the cluster and S1 spectra from each other. This fitting
method was developed by Uno et al. \cite*{uno99} and also presented in
Mitsuda et al. \cite*{mitsuda97}. We summarize below the flow of the
method and show the results. In this analysis, the SIS-0 and -1 are
added together and the 0.5--8.5 keV energy band is used.

First we defined a rectangular region of 10\arcmin\ length and
2\arcmin\ width as shown in Figure \ref{sis_image}, in which the
cluster and S1 are located along the major axis of the rectangle. We
set the position of the X-ray peak of the cluster in the SIS image at
the origin of the axes. We then made a 1-dimensional intensity profile
as a function of the position along the major axis by integrating the
photons along the minor axis.  In order to fit the observed projected
profile, we constructed a model function consisting of three
components: a cluster, point source, and background. For the cluster
component, we generated a projected cluster surface brightness profile
by convolving an image which has our best-fit $\beta$-profile (as
determined from the HRI image) with the PSF of the ASCA XRT and
integrating along the minor axis in the same manner as above. Then for
the point source, we derived a projected Point Spread Function by
integrating the Point Spread Function also in the same way. We assumed
the background flux to be constant along the major axis.

\begin{figure}
\psfig{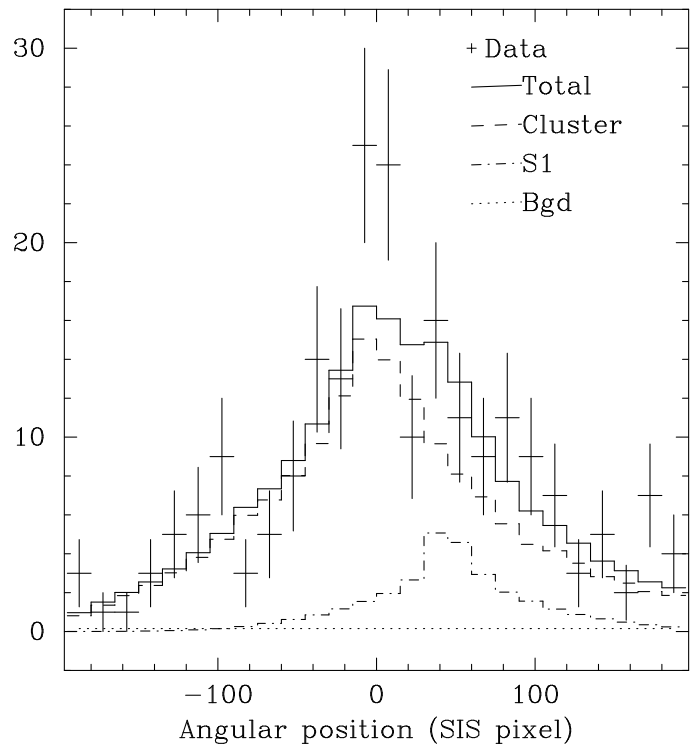}
\caption{Example of 1-dimensional image fitting. The crosses denote
the observed 1-dimensional profile of SIS in the 1.14--1.40 keV band
and the step function shows the best-fit model function(solid line). The three
model components, i.e. the cluster, Seyfert and background are
shown with dashed, dot-dashed, and dotted lines, respectively. 1 SIS
pixel corresponds to 0.026 arcmin.}
\label{imagefit}
\end{figure}

We fitted the observed 1-dimensional profile of seven energy bands
with the model derived above. The width of each energy band was
adjusted so that we can perform the image fitting with reasonable
statistics. Thus we employed seven bands: 0.50--0.90, 0.90--1.14,
1.14--1.40, 1.40--1.78, 1.78--2.61, 2.61--4.00 and 4.00-8.49 keV. In
Figure \ref{imagefit}, we show an example of the 1-dimensional
fit. The centroid of the projected $\beta$-profile was fixed at the
origin and the angular distance from the cluster to the point source
was at the value expected from the ROSAT HRI image i.e. 1.2
arcmin. The best-fit normalizations give fluxes of Cl0024+17 and S1 in
each energy bin.

\begin{figure}
\psfig{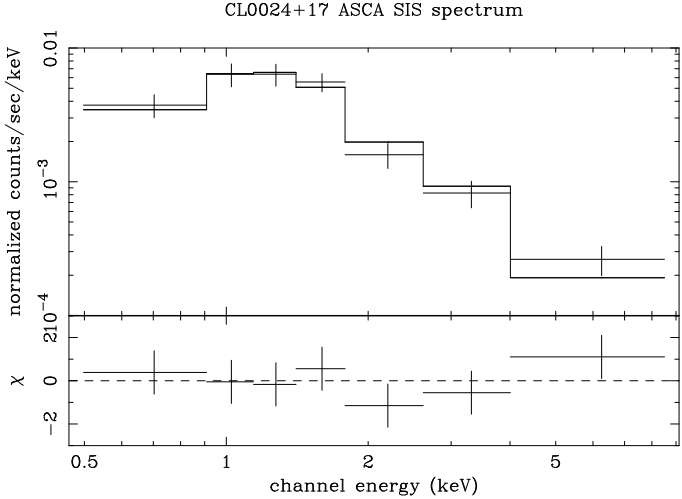}
\caption[]{X-ray spectrum of Cl0024+17 obtained by the 1-dimensional
image fit. The crosses denote the derived spectrum and the step
function shows the best-fit model function convolved with the X-ray
telescope and the detector response functions.}
\label{sis_spec}
\end{figure}

Then combining these flux values in multiple energy bins, we obtained
the spectrum of both Cl0024+17 and S1. We fitted the contamination-free
cluster spectrum with a thin-thermal Raymond-Smith model (Figure
\ref{sis_spec}). The neutral absorption was fixed at the galactic
value; $N_H = 4.2\times 10^{20} {\rm cm^{-2}}$. We fixed the metal
abundance at 0.3 solar. The chi-squared value is 3.3 for 5 degrees of
freedom. The X-ray temperature $kT$ is determined to be 5.7 keV with
a 90\% error range of 3.6 -- 10.6 keV.

\subsection{Total luminosity of the cluster} 
The cluster luminosity can easily be estimated from the total counts
measured on the ROSAT HRI image. In this image, we estimate a count
rate of $6.6 \times 10^{-3}$ cts s$^{-1}$ (or a total number of 765
counts) for the cluster emission only. This value is evaluated
directly from the cluster model issued from the fit, with no
contribution from S1 or S2. It is probably underestimated as we had to
stop the integration at a radius of about 900 $h_{50}^{-1}$ kpc, when
the signal could not be distinguished from noise. 
Assuming again a hydrogen column density of $N_H = 4.2 \times 10^{20}$
cm$^{-2}$ for the cluster and a temperature of 5.7 keV for a
Raymond-Smith plasma, we deduce a measured flux of $3.5 \times
10^{-13}$ erg s$^{-1}$ cm$^{-2}$ and a luminosity in the [0.1--2.4]
keV ROSAT band of $2.7 \times 10^{44} \ h_{50}^{-2}$ erg s$^{-1}$.
This value is quite similar to previous measurements in this cluster
from Einstein data \cite{henry82} and consistent with the measurement
given in B\"ohringer et al.  \cite*{bohringer99} but is clean from the
contribution of the contaminants S1 and S2. Note however that the
X-ray emission in Cl0024+17 is relatively weak when compared to other
cluster lenses \cite{fort94,allen98}. It is also satisfying to note
that this X-ray luminosity is not very different from the expected
value computed from the $L_X - T_X$ relation
\cite{markevitch98}. Indeed, Markevitch (1998) discusses the scatter
in the $L_X - T_X$ relation and finds not more than 20--30 \%\ in $T$
for a given $L_X$. Thus for the value quoted here for Cl0024+17, we
find a temperature range $T_X = 3.6^{+1.3}_{-1.0}$ keV, compatible
with the low side of the spectroscopically measured value. 

From the results of the previous fit and calculations, we also find a
central electron density of $n_0 \simeq 8 \times 10^{-3}$
cm$^{-3}$. The resulting cooling time is about $t_{cool} \simeq 7-9
\times 10^{9}$ yr, of the same order as the cluster age at the
redshift of $z=0.39$, in particular if we take into account a redshift
of formation $z_{\rm for} =2$ for the cluster ($t_{\rm cluster} \simeq
8 \times 10^9$ yr).  So no cooling flow signature is expected in the
X-ray signal, whatever the past merging history of the cluster.

\section{Mass analysis}
\subsection{X-ray mass determination}
The mass determination issued from the X-ray analysis was done in a
classical way, assuming that the $\beta$-profile also holds for the
gas density, and that the gas is in hydrostatic equilibrium. Cl0024+17
is a distant cluster, so no radial temperature distribution is
available. We will then consider the case of an isothermal cluster,
with $kT=5.7 ^{+4.9}_{-2.1}$ keV. 
At large distance from the center,
the effect of the core radius vanishes and the main uncertainty in the
absolute numbers is essentially related to the temperature
uncertainty, and then to the underestimation of the $\beta$
parameter. Under these assumptions, the total mass inside a radius $r$
writes as:
\begin{eqnarray*}
M(<r) & = & \frac{3 \beta k T r_c}{\mu m G} \ \frac{(r/r_c)^3}
{1 + (r/r_c)^2} \\
 & \simeq & \left( 1.9 ^{+1.6}_{-0.7} \right) \times 10^{13} \ 
\frac{(r/r_c)^3}{1 + (r/r_c)^2} \ \ h_{50}^{-1} \ M_\odot \\
\end{eqnarray*}
and the 2D-projected mass, integrated within a cylinder of radius $R$ 
writes as: 
\begin{eqnarray*}
M_{\rm 2D} (<R) & = & \frac{3 \beta k T r_c}{\mu m G} \ 
\frac{\pi}{2} \ \frac{(R/r_c)^2}{\sqrt{1 + (R/r_c)^2}} \\
 & \simeq & \left( 3.0 ^{+2.6}_{-1.1} \right) \times 10^{13} \ 
\frac{(R/r_c)^2}{\sqrt{1 + (R/r_c)^2}} \ \ h_{50}^{-1} \ M_\odot \\ 
\end{eqnarray*}
The integration inside a radius of 1 $h_{50}^{-1}$ Mpc gives a total
mass for the cluster
\[ M_{\rm tot} (r<1 h_{50}^{-1} {\rm Mpc}) = \left( 2.9 ^{+2.5}_{-1.1} 
\right) \times 10^{14} \ h_{50}^{-1} \ M_\odot \] 
and a projected mass inside a cylinder of radius 1 $h_{50}^{-1}$ Mpc
\[ M_{\rm 2D,tot} (R<1 h_{50}^{-1} {\rm Mpc}) = \left( 4.5 ^{+3.9}_{-1.6} 
\right) \times 10^{14} \ h_{50}^{-1} \ M_\odot \]

\subsection{Cl0024+17 as a gravitational lens} 
Two mass estimates were proposed for Cl0024+17 from the lensing
analysis. The first one comes from the HST observations of the giant
arc system and its modeling \cite{kassiola92,smail96}.  The mass
distribution (projected dark matter) in the center is very peaked with
a small core radius ($\sim 40 h_{50}^{-1}$ kpc) and a small
ellipticity. The projected mass inside the radius of the arc is
\[ M_{\rm proj} (R < R_{\rm arc}) = (2.0 \pm 0.2) \times  10^{14} 
\ h_{50}^{-1} \ M_\odot \] 
with $R_{\rm arc} = 35'' = 220 \ h_{50}^{-1}$ kpc.  
An updated value of this mass can be found in Broadhurst et
al. \cite*{broadhurst99} with the measurement of the arc redshift 
($z_S = 1.675$): 
\[ M_{\rm proj} (R < R_{\rm arc}) = (2.6 \pm 0.06) \times  10^{14} 
\ h_{50}^{-1} \ M_\odot \] 
It compares with the X-ray mass deduced from our analysis:
\[ M_{\rm 2D,X} (R < R_{\rm arc}) = \left( 0.96 ^{+0.82}_{-0.35} \right) 
\times 10^{14} \ h_{50}^{-1} \ M_\odot \]
The weak lensing mass was detected by Bonnet et
al. \cite*{bonnet94} in their pioneering work on the weak shear
detection at large distance from the cluster center. Although their
mass inversion was rather simple, they were able to constrain the
surface mass density profile and to estimate the projected mass of the
cluster inside a radius of 3 $h_{50}^{-1}$ Mpc, considering the mass
profile of a singular isothermal sphere: 
\[ M_{\rm proj} (R < 3 h_{50}^{-1} {\rm Mpc}) \simeq 4 \times 10^{15} 
\ h_{50}^{-1} \ M_\odot \] 
Changing the profile from a de Vaucouleurs law to a power law (close
to isothermal) induces a mass range from $(2.4 - 4) \times 10^{15} \
h_{50}^{-1} \ M_\odot$. 
Further analysis is in progress to better invert this mass profile,
using the inversion procedures developed by Schneider \& Seitz
\cite*{schneider95} and Seitz et al. \cite*{seitz95,seitz98}.
Again we can compare with the extrapolated total mass deduced from the 
X-ray profile: 
\[ M_{\rm 2D,tot} (r < 3 h_{50}^{-1} {\rm Mpc}) = \left( 1.4 
^{+1.2}_{-0.5} \right) \times 10^{15} \ h_{50}^{-1} \ M_\odot \] 
In both cases there is a discrepancy between the mass estimates, from a
marginal value up to a factor 3. This result is similar to what has
already been found in many clusters such as Abell 2218 and Abell 1689
\cite{miralda95,squires96} and is discussed below.

\section{Resolving the mass discrepancy ?}  
The combination of the different mass estimates presented here in a
homogeneous scaling and with error bars leads to a typical mass
discrepancy of a factor 1 to 3. We may ask how to solve this apparent
paradox, in particular by looking in more detail through each set of
data and through the methods used to derive mass measurements. First,
concerning X-ray data, it is now well known that HRI data are limited
by the background level and that the extrapolation of the X-ray
profile underestimates the $\beta$-value or equivalently the
logarithmic slope of the profile. This error on $\beta$ can lead to an
underestimate of the total mass by a typical factor of 1.5
\cite{bartlemann96}. This effect is particularly important in the
case of Cl0024+17 which is a distant cluster. So, due to detection
limits, the X-ray flux is not measured at a radius large enough with
respect to the core radius, and the X-ray mass is significantly
underestimated. The extrapolation of the X-ray profile well above the
detection limit is also questionable and we cannot exclude a steeper
profile at large distance.  The uncertainty on the temperature measure
is a large source of error if we strictly apply the $\chi ^2$
statistics. But with a-priori information coming from the well-known
$L_X-T_X$ relation, it may be difficult to increase this measure by a
factor of two necessary to solve the discrepancy. The satisfying
consistency between the measured values of the temperature and the
X-ray luminosity, coming from independant X-ray data is important and
gives confidence in the X-ray mass determination of the cluster.

The shear mass is questionable for several reasons: first the shear
measurement could be improved. In particular, better procedures
correcting instrumental defects such as PSF anisotropy, instrumental
flexure or atmospheric refraction have recently been developed
\cite{kaiser95,waerbeke97,mellier99} and mass reconstructions are
becoming more reliable. Second, the mass measured inside a radius of
$3 h_{50}^{-1}$ Mpc is strongly dependent on the slope of the
potential at large distance. We
can estimate that an uncertainty of a factor 2 remains in the mass
derived from weak lensing measurements, the preferred value of a
singular isothermal profile being an upper limit. Since the pioneering
work of Bonnet et al. \cite*{bonnet94} Navarro, Frenk \& White
\cite*{navarro96} have shown from N-body/hydrodynamical simulations
that massive dark matter halos may follow a ``universal''
profile. This profile has not been applied to the cluster Cl0024+17
yet, and could decrease the total mass at large distance with respect
to the isothermal distribution.  But whatever the exact dark matter profile
shape (NFW or De Vaucouleurs), the assumption of an isothermal 
distribution of the gas cannot be applicable anymore. 
In that case, the mass reconstruction from the X-ray profile must 
be refined. A possibility should be to include a $\beta$-model with a
polytropic index $\gamma$ representative of the departure from
isothermality \cite{cavaliere78}. With values of $\gamma$ in the range 
$[1,1.2]$, such a model can be consistent with a steep potential like
the NFW one and may reduce the discrepancy between the mass
estimates. In
addition, the $\beta$-profile is mainly used as a practical
parametrization to fit the X-ray surface brightness with a simple
analytical de-projection. The main danger is that we fit only part of
the profile seen, and then extrapolate to larger radii to reach the
weak shear measures of Bonnet et al. \cite*{bonnet94}. We can justify
this by observing that for clusters at low redshift where
the X-ray emission can be traced at larger radii, a relatively
straight power law slope is generally observed. But this approach is
still not free from systematic errors, as also commented above. 

Another point to mention is the
redshift distribution of the sources. Bonnet et al. \cite*{bonnet94}
simply assumed that all the sources were located in a single plane of
background sources, with $z_S \sim 0.8 - 1.2$, depending on the
potential. Changing the source distribution from $z_S \sim 0.8$ to
$z_S \sim 2$ would lower the lensing mass estimate by about 40
\%. Recently, Athreya et al. \cite*{athreya99} tried to give a correct
scaling of their mass reconstruction of the weak lensing map in the
cluster MS1008--1224 by using photometric redshift estimates of the
background sources. Using deep multi-color VLT data, they were able to
introduce a realistic redshift distribution, even for the very faint
sources for which we presently do not have spectroscopically measured
redshift distributions.  A detailed comparison between all the mass
determinations in this cluster \cite{carlberg96,lewis99} gave a
X-ray/lensing discrepancy of a factor 2 in the central region of the
cluster, while in the outer parts and on a larger scale, the agreement
is much better. The agreement with the dynamical M/L ratio was also
found quite satisfying. Finally, one should note that the weak lensing
analysis of Cl0024+17 was only done in one sector of the distant
regions of the cluster. It then may be subject to inhomogeneities in
the background distribution of the sources or in the mass fluctuations
on the line of sight. In the case of MS1008--1224, this is clearly
the case in one of the quadrants of the image \cite{athreya99}!

Last, we can also question the validity of the velocity dispersion
measured by Schneider et al. \cite*{schneider86} as it corresponds to
38 cluster members only. From a deep and wide field spectroscopic
survey aimed at studying in detail the dynamics of the galaxies in
this cluster, preliminary results show a bi-modal redshift
distribution of the galaxies \cite{czoske99} with no clear spatial
separation between the two modes. The main component of this
distribution corresponds to a velocity dispersion of 900 km s$^{-1}$
only, which decreases the virial mass by a factor of 2.  This also
reconciles the virial mass with the X-ray mass, and the cluster
characteristics then match the $\sigma -T_X$ correlation. In addition,
this confirms the fact that Cl0024+17 has a complex dynamical
structure, and that the X-ray gas may not be in a fully hydrostatic
state. Departures from an hydrostatic equilibrium possibly traced 
by the elliptical shape of the X-ray emission or by the dynamics of
the galaxies in the field of the cluster may fix some of the
discrepancy in the different mass estimates. 
More details on the galaxy dynamics and the shape of the
potential will be discussed in a forthcoming paper \cite{czoske00},
as well as the effects of the projection of several mass
components on the different mass estimators.

\section{Conclusion}
The detailed analysis of both spatial and spectral X-ray data
presented in this paper leads to several results. First and as one of
the initial goals of this work, we marginally detect a weak X-ray
counter-part of the shear perturbation identified by Bonnet et al.
\cite*{bonnet94}. This definitely needs to be confirmed by deeper
X-ray data, waiting for the new generation of X-ray satellites (XMM
and Chandra) for a better sensitivity. In addition our results are
quite similar to what was found by Erben et al. \cite*{erben99}. We
may be close to resolve the nature of the so-called ``dark clumps''
which begin to be detected in wide field shear maps around clusters.
They may correspond to either small but massive groups in the
vicinity of rich clusters of galaxies or distant clusters, too faint
in the optical to be associated with the shear perturbations. In both
cases, hot gas seems to be present, although in small
quantities. Further studies and a better statistics on these ``dark
clumps'' are required to better understand the cosmological
consequences.

From our deep ROSAT HRI image, one of the deepest fields observed with
this instrument, we were also able to derive number counts of X-ray
sources at levels as faint as $\sim 10^{-14}$ erg s$^{-1}$ cm$^{-2}$,
quite compatible with number counts observed in other fields with the
PSPC instrument. Only one optical identification is presently
available and corresponds to QSO. We suggest that the main component
of the faint X-ray sources are AGNs \cite{zamorani99}.

Finally, concerning the cluster itself, we were able to carefully
examine its 2D X-ray properties. A temperature measurement was
produced, clear from the contaminating sources identified on the
image. Although still rather uncertain, this temperature is quite
compatible with the total X-ray luminosity of the cluster, giving good
internal consistency of the X-ray data. On the contrary, an
interesting mass discrepancy arises when one compares the mass deduced
from the X-ray analysis with the lensing mass, either in the center of
the cluster (strong lensing) or in the outer parts of the cluster
(weak lensing). This apparent discrepancy is even stronger than in
many other clusters where it has been pointed out
\cite{miralda95,squires96}. But it can be significantly reduced by a
very careful analysis of the overall data. Future progress in the
study of the mass distribution in clusters will be boosted soon due to
the new generation of X-ray satellites, better understanding of the
weak shear measurements in deep wide field images and the multi-object
spectroscopic facilities available on 8 meter-class
telescopes. Joining these efforts together will allow to learn much
more on the dynamics and mass contents in clusters of galaxies.

\acknowledgements We wish to thank A. Blanchard, J. Bartlett and
J.-P. Kneib for fruitful discussions. This work was supported by the
Centre National de la Recherche Scientifique through the Programme
National de Cosmologie and by a European TMR
network programme: ``Gravitational Lensing: New Constraints on
Cosmology and the Distribution of Dark Matter'' under contract
No. ER-BFM-RX-CT97-0172 from the European Commission ({\tt
http://www.ast.cam.ac.uk/IoA/lensnet/}). O. Czoske thanks the EC for
financial support under contract No. ER-BFM-BI-CT97-2471.

%

\end{document}